\documentclass[%
 reprint,
 amsmath,amssymb,
 aps,
floatfix,
]{revtex4-2}
\usepackage{multirow}
\usepackage{algpseudocode}
\usepackage{algorithm}
\usepackage{hyperref}
\usepackage{placeins}
\usepackage{color}
\usepackage{graphicx}
\usepackage{dcolumn}
\usepackage{bm}

\graphicspath{{images/}}

\begin{document}

\preprint{APS/123-QED}

\title{Critical dynamics of long range models on Dynamical L\'evy Lattices}

\author{Riccardo Aiudi}
 \affiliation{Dipartimento di Scienze Matematiche, Fisiche e Informatiche,
Universit\`a degli Studi di Parma, Parco Area delle Scienze, 7/A 43124 Parma, Italy}
\affiliation{INFN, Gruppo Collegato di Parma, Parco Area delle Scienze 7/A, 43124 Parma, Italy}

\author{Raffaella Burioni}
 \affiliation{Dipartimento di Scienze Matematiche, Fisiche e Informatiche,
Universit\`a degli Studi di Parma, Parco Area delle Scienze, 7/A 43124 Parma, Italy}
\affiliation{INFN, Gruppo Collegato di Parma, Parco Area delle Scienze 7/A, 43124 Parma, Italy}

\author{Alessandro Vezzani}
\affiliation{Istituto dei Materiali per l'Elettronica ed il Magnetismo (IMEM-CNR), Parco Area delle Scienze, 37/A-43124 Parma, Italy}
 \affiliation{Dipartimento di Scienze Matematiche, Fisiche e Informatiche,
Universit\`a degli Studi di Parma, Parco Area delle Scienze, 7/A 43124 Parma, Italy}
\affiliation{INFN, Gruppo Collegato di Parma, Parco Area delle Scienze 7/A, 43124 Parma, Italy}
\date{\today}

\begin{abstract}
We investigate critical equilibrium and out of equilibrium properties of a ferromagnetic Ising 
model in one and two dimension in the presence of long range interactions, $J_{ij}\propto r^{-(d+\sigma)}$. 
We implement a novel local dynamics on a \emph{dynamical}  L\'evy lattice, that correctly reproduces the static critical exponents known in the literature, as a function of the interaction parameter $\sigma$. Due to its locality the algorithm can be applied to investigate dynamical properties, of both discrete and continuous long range models. We consider the relaxation time at the critical temperature and we measure the dynamical exponent $z$ as a function of the decay parameter $\sigma$, highlighting that the onset of short range regime for the dynamical critical properties appears to occur at a value of $\sigma$ which differs from the equilibrium one. 
\end{abstract}

\maketitle


\section{\label{sec:level1}Introduction\protect}

Systems with non-local long range interactions are known to give rise to interesting physics in classical \cite{campa_physics_2014} and quantum \cite{defenu_long_2021} many-body
regimes, both at equilibrium and out-of equilibrium.  Long range interactions can induce spontaneous symmetry breaking even at low dimensions \cite{dyson_1969} and, at continuous phase transitions, a sufficiently slow decay can modify the universality classes, resulting in critical exponents which depends on the interaction decay \cite{fisher_critical_1972}. Out of equilibrium, long range interactions modify  the dynamical exponents in coarsening phenomena, \cite{Bray1,Bray2,christiansen_phase_2019,christiansen_aging_2020},  also giving rise to non trivial metastable states that can affect the dynamics  \cite{Corberi1,Corberi2}.

A wide interest in the field has been devoted to the investigation of magnetic models on lattices in the presence of long range interactions of the type $J_{ij}\propto r^{-(d+\sigma)}$. The accepted description of the critical properties in the ferromagnetic case was given by Sak \cite{sak_recursion_1973}. For $\sigma<d/2$,  the critical behavior is mean field, while in the so called "long range" regime, $d/2<\sigma< 2- \eta_{SR}$, the critical exponents depend non-trivially on $\sigma$ ($\eta_{SR}$ is the exponent of the critical correlation function in the short range model). For large enough $\sigma>2- \eta_{SR}$, the short range behavior is recovered.  

In this class of models, very often studies are based on numerical approaches. 
In  \emph{physical} long range models, interactions involve all degrees of freedoms since all pairs of sites interact and thus they form a fully connected model on a complete graph with weighted links. This implies that the number of interactions scales as $O(N^2)$, requiring large resources for simulations. 
In recent years, several solutions to this problem have been proposed. Cluster algorithms \cite{luijten_monte_1995,fukui_order-n_2009} can simulate long range model with a computational cost of $O(N\log{N})$ or $O(N)$ and reduce the relaxation time, so that they are not influenced by critical slowing down, at the price of a non local dynamical evolution. On the other hand the kinetic Monte Carlo 
\cite{christiansen_phase_2019,christiansen_aging_2020,Corberi1,Corberi2} provides an effective tools for simulating a local dynamics only at very low temperatures.

Another interesting solution are L\'evy lattices \cite{leuzzi_dilute_2008,Leuzzi_diluted2,Katzgraber,berganza_critical_2013}. These are random diluted graphs with interactions between pairs which are constant and occur with probability $\propto r^{-(d+\sigma)}$, with the total number of interactions being $O(N)$. L\'evy lattices drastically reduce the computational cost while keeping a local dynamics, however an average over different realizations is required. In $1$ dimension, L\'evy lattices seem to fall in a different universality class than their fully connected counterparts, due to long range correlations induced by the disorder in the random lattice realization \cite{defenu_1d}, while in $2$ dimension consistent results have been obtained in the XY model \cite{berganza_critical_2013,cescatti}.

In this paper, we introduce an alternative local dynamics for the long range model based on a \emph{dynamical} L\'evy lattice, {that is a dynamical sampling at each time step
of the long range interacting model, in the spirit of the
q-Ising model \cite{jedrzejewski_oscillating_2015,chmiel_q-neighbor_2018}. In a nutshell, each spin interacts with a constant interaction J
only with q neighbours, randomly drawn from the corresponding long range probability distribution.}
Unlike L\'evy lattices, the underlying graph is not fixed before the dynamics take place, but evolves dynamically with the system and it is built during the simulation. The system can be thought of as living on a temporal L\'evy graph \cite{holme_temporal_2012,ubaldi_asymptotic_2016,ubaldi_burstiness_2017}, in which at each time step the underlying structure evolves and it is reshuffled.
{As a result, the long range correlations characterizing the single disordered realization of static L\'evy lattices are eliminated. Such correlations are known to deeply affect the behavior of the system, at least in low dimension as shown in \cite{defenu_1d}. Moreover, in the dynamical L\'evy lattice physical quantities are computed by averaging over the full dynamic evolution of the model and thus feature the same symmetries of the statistical model on the fully connected graph. This is expected to provide a significant numerical advantage with respect to simulations on the static L\'evy lattice in particular at small $\sigma$, which are affected by large fluctuations in different quenched realizations at finite size \cite{cescatti}.}

On the other hand, in the dynamical L\'evy lattices detailed balance is not guaranteed, so we need to compare our results with that of static fully connected lattices to show that they belong to the same universality class. Therefore, we first test the validity of the new algorithm by studying numerically the critical behaviour of $1$ and $2$-dimensional Ising model in the presence of long range interactions, comparing the results with the Sak's scenario and with state-of-the-art simulations \cite{luijten_finite-size_1996,luijten_classical_1997,luijten_boundary_2002,picco_critical_2012,angelini_relations_2014}.  We then present a novel numerical measure of the dynamical exponent $z$ at the critical temperature, which governs the behaviour of the relaxation time at criticality.  {This is a critical dynamical quantity that cannot be measured with the non-local cluster algorithm \cite{luijten_monte_1995,fukui_order-n_2009} nor with the
kinetic Monte Carlo \cite{christiansen_phase_2019,christiansen_aging_2020}}.
We consider the full interesting range of $\sigma$. In the mean field regime $\sigma<d/2$, we show that $z=d/2$, consistently with the relaxation properties of a Curie Weiss model. At $\sigma=d/2$, the super-diffusive behavior $z=\sigma$, characterising a free random walk with long range motion, is recovered.
At larger $\sigma$ interaction starts to play a role, giving rise to perturbative contributions and, interestingly, the transition to the short range behavior seems to occur at a value of $\sigma$ larger than that for the short range regime at the equilibrium.   

The paper is organized as follows: in section \ref{sec:model} we briefly describe the long range model and we summarize the existing algorithms with their features and limitations. In section \ref{sec:dll} we present our dynamics and in section \ref{sec:eta} we numerically test its validity comparing the results with the Sak's prediction. In section \ref{sec:zeta} we present our results for the dynamical critical exponent $z$. Finally, in section \ref{sec:conclusion} we summarize and discuss future perspectives of our work. 

\section{\label{sec:model}The long range Ising Model}
The ferromagnetic Ising model with long range interaction on a hypercubic lattice of dimension $d$ is described by the Hamiltonian:
\begin{equation}
\label{hamiltonian}
    H = -\frac{1}{2}\sum_{i,j=1}^N J_{ij}\sigma_i\sigma_j,
\end{equation}
where $\sigma_i$ are the usual Ising spins and $J_{ij}\propto r_{ij}^{-(d+\sigma)}$, with $r_{ij}$ the euclidean distance between $i$ and $j$,  $\sigma$ being the parameter for the decay of the long range interaction. The interactions have here infinite range and connect all the spin, i.e. the system is fully connected.
According to Renormalization Group calculations \cite{fisher_critical_1972,sak_recursion_1973,defenuRG}, the model exhibits three different behaviours depending on the value of $\sigma$:
\begin{itemize}
    \item $\sigma \in (0,\frac{d}{2})$: the system is in a mean field (MF) regime;
    \item $\sigma \in (\frac{d}{2}, 2-\eta_{SR})$, where $\eta_{SR}$ is the critical exponent $\eta$ characterizing the decay of the correlation function at criticality in the short range model in $d$ dimension: the system belongs to a universality class different from the MF one and the critical exponents depend on the value of $\sigma$. This region will be referred as the long range (LR) region;
    \item $\sigma > 2-\eta_{SR}$: the interaction decays so fast that the system becomes equivalent to the corresponding short range (SR) model;
\end{itemize}
The boundary $\sigma = 2-\eta_{SR}$ between LR and SR regions is indeed the main point of Sak's prediction. 
In $1$ dimension the LR phase is observed for $1/2<\sigma<1$, while for $\sigma>1$ the system does not present a critical transition at finite temperature, and at $\sigma=1$ a Kosterlitz-Thouless transition occurs \cite{Cardy_1981} with non trivial features \cite{Bar_2014,Barma_fluctuation_2019}.

\subsection*{\label{sec:algos}Numerical approaches to long range models}

As mentioned above, the simulation of the fully connected model is computationally costly, and several algorithms and  techniques have been introduced to study long range models numerically. We now briefly discuss the approaches introduced so far.

{\it Cluster Algorithms.}
Following the idea of Swendsen-Wang and Wolff \cite{swendsen_nonuniversal_1987,wolff_collective_1989}, cluster algorithms have been designed to simulate long range Ising models with a computational cost of $O(N)$ \cite{luijten_monte_1995,fukui_order-n_2009}. The basic step of these algorithms is to flip arbitrary large clusters of spins by preserving the detailed balance and reducing the relaxation time.
Indeed large clusters are updated in a single step and uncorrelated configurations are rapidly obtained even at the critical point. This avoids the critical slowing down, which is the typical bottleneck in terms of computational cost. 
This algorithm has been used to test the Sak's scenario and to confirm that it gives the correct description of the critical behavior in the Ising case \cite{luijten_finite-size_1996,luijten_classical_1997,luijten_boundary_2002,fukui_order-n_2009,angelini_relations_2014}.
However, as there is no local dynamics, the algorithm cannot be used to study the temporal evolution. Moreover, identification of clusters is straightforward for the Ising model while it cannot be implemented in models with continuous variables. {Recently, a cluster algorithm has also been introduced to study long-range percolation \cite{gori}. }

{\it The Kinetic Monte Carlo Algorithm.}
Recently an implementation of the Kinetic Metropolis algorithm has been introduced  
for the long range Ising model \cite{christiansen_phase_2019,christiansen_aging_2020,Corberi1,Corberi2}. In this local algorithm, the time intervals in which no spin flip occurs are directly estimated after each move and the rejection rate vanishes. The implementation turns out to be very efficient at low temperatures, when spin flips are very unlikely, while at higher temperatures (e.g. at criticality)  the efficiency is comparable to standard Metropolis. 
The  Kinetic Monte Carlo Algorithm has been applied to study non-equilibrium properties in coarsening dynamics, that is the evolution of the system quenched from $T=\infty$ to a temperature lower than the critical one \cite{Bray1,Bray2}.

{\it L\'evy lattices.}
The L\'evy lattice is a diluted graph with adjacency matrix $A_{i,j}=0,1$, built to display the same properties of the long range fully connected model \cite{leuzzi_dilute_2008,Leuzzi_diluted2,Katzgraber,berganza_critical_2013}.
In particular, the long range ferromagnetic model on a $d$-dimensional lattice is approximated by a graph in which two sites are connected with a probability proportional to the long range interaction, i.e. $A_{ij} = 1$ with probability $P_{ij} \propto r_{ij}^{-(d+\sigma)}$. On the diluted L\'evy lattice, the usual Monte Carlo simulations are implemented with saving of computational cost, since each node is connected to a finite number of edges.
There are no analytical proofs that models on L\'evy lattices are equivalent to their long range fully connected counterparts. In particular, for $1$-dimensional lattices, numerical evidences suggest that static long range correlations can modify the critical properties in the free model, changing the spectral dimension \cite{defenu_1d}. On the other hand, in the $2$-dimensional case, the  spectral dimension of the L\'evy lattice seems to coincide with that of the long range, suggesting that the difference is limited to low dimensions \cite{berganza_critical_2013,bighin_universal_2022}. At difference with the cluster algorithm, the L\'evy diluted lattice can be used to simulate continuous symmetry models. Consistent simulations have been performed for the $2$-dimensional $XY$ model with long range interaction \cite{berganza_critical_2013,cescatti}, in the presence of a Kosterlitz Thouless phase transition \cite{giachetti_berezinskii-kosterlitz-thouless_2021}. 
On finite systems, simulations on L\'evy lattices strongly depend on the graph realization and averages over a large number of samples are necessary to obtain stable results. In this context the problem of self averaging is still an open issue; moreover the averaging procedure can be very demanding from a computational point of view \cite{cescatti}.

\section{\label{sec:dll}The Dynamical L\'evy lattice }

The starting point of our approach is the q-Ising model \cite{jedrzejewski_oscillating_2015,chmiel_q-neighbor_2018}. In this algorithm, at each time step a spin $\sigma_{i}$ is chosen randomly and it interacts with the field produced by $q$ neighbors, which are also randomly drawn uniformly among the remaining $N-1$ spins. The spin $\sigma_{i}$ flips according to a Metropolis \cite{metropolis_monte_1949} or Glauber \cite{glauber_timedependent_1963} prescription. 
As shown in \cite{park_tricritical_2017} this model has two fluctuating variables, the spins and the links. The former are in contact with the heat bath, while the links are randomly rewired during the dynamics, without any acceptance-rejection procedure. The links can be considered as being in thermal contact with a heat bath of temperature $T=\infty$, which implies rewiring  with probability $1$. With two different heat baths governing the dynamics, detailed balance is not satisfied and thermal equilibrium is not obvious. 

{
Our idea is to simulate a long range $q$-Ising model, which we expect to belong to the same universality class as the long range Ising model.  In the \textit{Dynamical L\'evy lattice} (DLL), first we select the spin $\sigma_i$ and thereafter its $q$ neighbors are randomly chosen from the nodes of the lattice, based on a power-law probability distribution $P_{ij}\propto r_{ij}^{-(d+\sigma)}$.  Then $\sigma_i$ flips, according to a constant interaction $J$ with the $q$ randomly selected spins. Note that in this way the decay of the interaction as a power law is recovered in a statistical sense. In particular, the model can be regarded as living on a \emph{dynamical} L\'evy lattice, because the graph has an adjacency matrix which evolves with time, i.e. $J_{ij}\rightarrow J_{ij}(t)= J\times A(i,j,t)$, where $A(i,j,t) = 1$ if at time $t$ the spin $\sigma_j$ is drawn as a neighbour of $\sigma_i$ and $0$ otherwise.

We observe that, by definition, random walks on a DLL exhibit the same behavior as random walks on the fully connected long-range graph: in this case, in fact, the walker jumps at each step from the starting site $i$ to any nodes of the network $j$ with probability $P_{ij}\propto r_{ij}^{-(d+\sigma)}$. 
This is exactly the same probability which is used on a DLL at each step to choose the $q$-neighbors of the walker, and then the jumps occur uniformly among these $q$ nodes. So the two dynamics coincide and, hence, the spectral dimension, as measured from the return probability of the random walker \cite{alexander_density_1982,Burioni_RW_2005}, are the same on the fully connected graph and on DLL. In contrast, on a L\'evy lattice the walker evolves on a random static network, thus in the presence of correlations. For example, when the walker crosses a link connecting two long-distance sites, it has a significant probability of going back along the same link that remains active on the static graph, while it fails to reach in a few steps the lattice sites that are not connected by long-distance links in that specific quenched realization. This induces static long-range correlations that are not present in either the fully connected lattice or the DLL. In particular, such correlations in $1$ dimension \cite{defenu_1d} are able to modify the spectral dimension with respect to the fully connected long range model.} 

{
Similarly, in the DLL the correlation functions $\langle \sigma_i \sigma_j \rangle$ are computed by averaging over the dynamic evolution of the model and thus naturally possess the symmetries of the statistical model on the fully connected graph. In contrast, on a L\'evy lattice, the $\langle \sigma_i \sigma_j \rangle$ depend on the quenched realization of the random structure and disorder breaks the original translation invariance. Therefore, the original spatial symmetries are recovered only after averaging over several realizations of the random lattice. We observe that this averaging procedure can also be very numerically demanding \cite{cescatti}.  
}

 The method we propose appears very flexible and can be applied to several statistical models both with discrete and continuous symmetry. 
Finally, the algorithm displays the same efficiency at any temperature even if, due to its local nature, we expect to observe the typical slowing down at criticality. An important point is that since on the DLL detailed balance is not valid and the equilibrium distribution is not known, the equivalence in terms of critical exponents with the fully connected long range system has to be checked.

\section{\label{sec:eta}Equilibrium critical properties}

Accordingly to Renormalization Group calculations, ferromagnetic long range models have critical exponents which depends on the parameter $\sigma$ \cite{fisher_critical_1972,sak_recursion_1973,defenuRG}. The three different regimes (MF, LR and SR) can be characterized by their critical exponents, for example the one governing the scaling of the magnetic susceptibility $\chi$, which can be defined as the magnetization $M$ fluctuations, $\chi \propto \langle M^2\rangle- \langle |M|\rangle^2$. We call this exponent $y$.

For $\sigma < \frac{d}{2}$ the  susceptibility exponent is equal to the MF one, $y = \frac{d}{2}$ \cite{luijten_classical_1997}; for $\frac{d}{2} <\sigma< 2-\eta_{SR} $ the system exhibits long range behaviour and the exponent is related to that of the spatial correlation $2-\eta$ and depends on $\sigma$: $y = 2-\eta = \sigma$; finally for  $\sigma > 2-\eta_{SR} $ the system is equivalent to the corresponding short range model, and so the exponent: $y = 2-\eta =2- \eta_{SR}$. The Sak's prediction for the boundary between LR and SR regions is widely considered valid and many numerical studies corroborate it \cite{luijten_finite-size_1996,luijten_classical_1997,luijten_boundary_2002,angelini_relations_2014}. Thus, one way to test the validity of the Dynamical L\'evy lattice is to measure this exponent and to compare it with the expected one.
It is known that strong finite size effects are present in long range models \cite{picco_critical_2012,angelini_relations_2014}.  One way to proceed is to include higher order terms in the scaling form of the thermodynamic functions {\cite{angelini_relations_2014}}. 


Close to criticality the scaling form of susceptibility is 
\begin{equation}
    \label{eq:scaling chi}
    \chi \sim L^{y}\cdot \tilde{\chi}(t L^{\frac{1}{\nu}}),
\end{equation}
where $L$ is the linear size of the system, $t = \frac{T-T_c}{T_c}$ the reduced temperature ($T_c$ is the critical temperature) and $\tilde{\chi}$ is a scaling function. In infinite systems $\chi$ diverges at $T_c$, but in finite ones there is a size-dependent temperature $T_c(L)$ where it has a maximum. According to Eq. \eqref{eq:scaling chi}, this maximum follows a power law behaviour $\chi_{max}(L)\sim L^y$
which can be efficiently used to extrapolate the exponent $y$ (see Appendix \ref{app:peaks}).

\subsection{$1$-Dimensional chain}
\begin{figure}
    \centering
    \includegraphics[scale = 0.8]{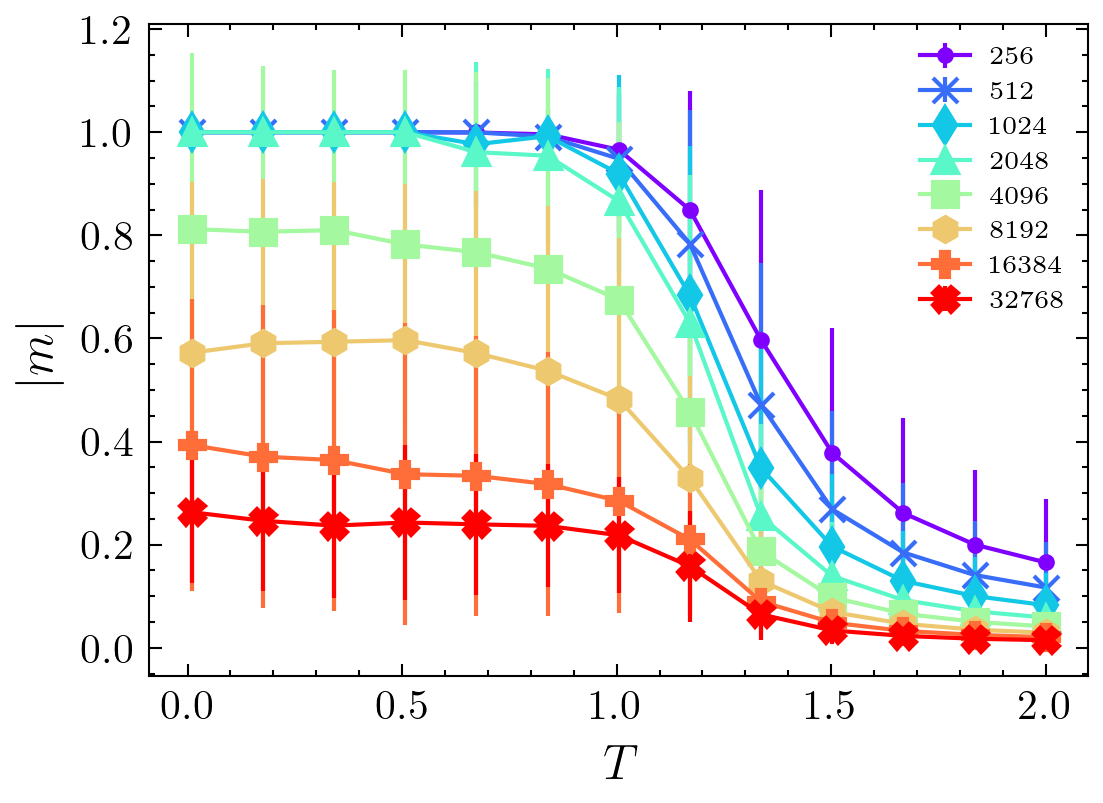}
    \caption{$|m|$ behaviour with the temperature as we increase the system size for $\sigma = 1.2$ ($1$-dimensional case).}
    \label{fig:120}
\end{figure}

\begin{figure}
    \centering
    \includegraphics[scale = 0.8]{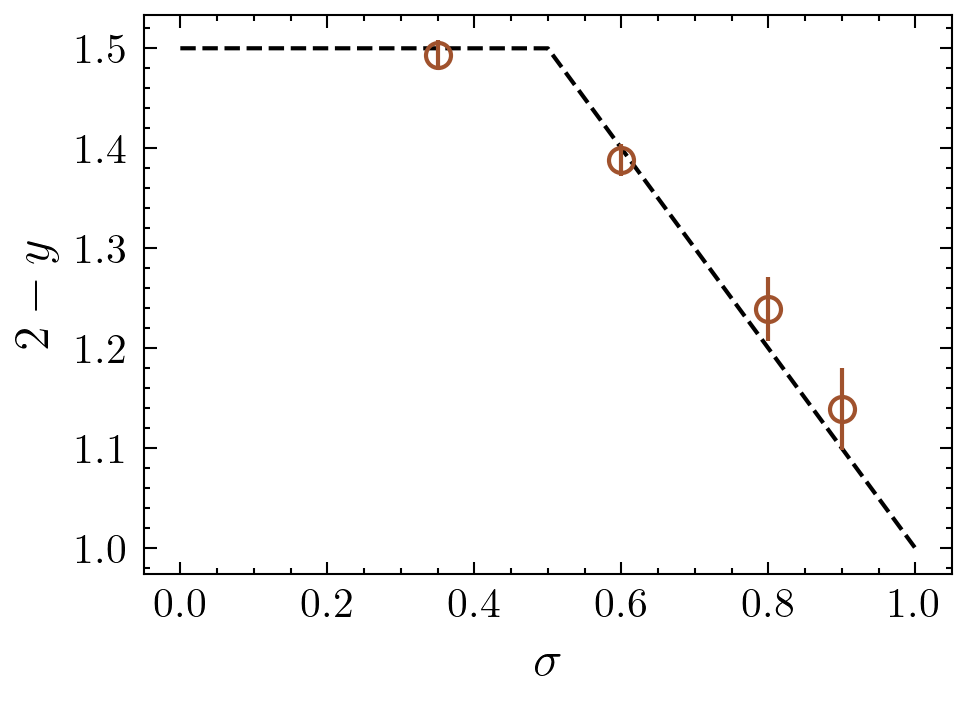}
    \caption{$2-y$ critical exponent extrapolated with our method, for the $1$-dimensional case. The black dashed line is the theoretical expectation. Numerical data are provided in Table \ref{tab:1d} in Appendix \ref{app:tables}.
    \label{fig:eta_1d}}
\end{figure}

We first focus on the $1$-dimensional spin chain (see Appendix \ref{app:algo} for details). We consider sizes from $L=2^8$ to $L=2^{15}$ and five values of the $\sigma$ parameter: $0.35$ for the MF region, $0.6$, $0.8$ and $0.9$ for the LR one and finally $1.2$ for the SR region, where the $1$-dimensional classical Ising model behaviour is recovered and the system does not undergo a phase transition. We check this behaviour by studying the absolute value of the magnetization per spin $m$, for different sizes. We find that $m$ goes to zero for each temperature as the size is increased, as it is clearly shown in Figure \ref{fig:120}. For the other values of $\sigma$ we find good agreement with theoretical expectations and we collect the results in Table \ref{tab:1d} and Figure \ref{fig:eta_1d}. {
An interesting regimes in the $1$-dimensional  case is observed when $\sigma$ approaches one and a Kosterlitz-Thouless transition occurs \cite{Cardy_1981}. However, in this limit we observe that simulations are quite demanding (see the large error on the exponents in Figure \ref{fig:eta_1d}) and a detailed study requires further investigations. 
}

\subsection{$2$-Dimensional lattice}
\begin{figure}
    \centering
    \includegraphics[scale = 0.9]{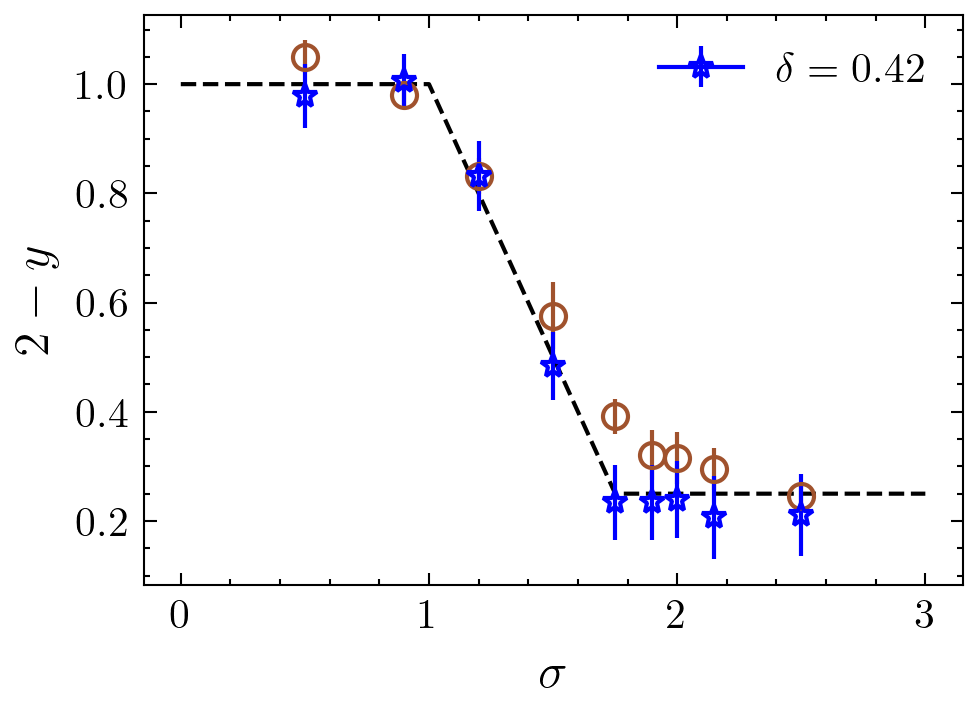}
    \caption{$2-y$ critical exponent extrapolated with our method, for the $2$-dimensional case. The black dashed line is the theoretical expectation. The brown circles are the results of our algorithm (DLL) and the blue star is the value found when using next to leading order corrections $(\delta = 0.42)$. Numerical data are reported in Tables \ref{tab:2d} and \ref{tab:2d_correction} in Appendix \ref{app:tables}}
    \label{fig:eta_2d}
\end{figure}

{
In the $2$-dimensional case we consider a wide range of $\sigma$'s in the MF, LR and SR regimes. The results for the critical exponent $\eta$ are summarized in Figure \ref{fig:eta_2d} and show a deviation from the theoretical prediction for value of $\sigma$ at the LR-SR boundary, where it is well known that there are large finite size effects as pointed out in \cite{picco_critical_2012,angelini_relations_2014} and the finite system is sensitive to boundary conditions (see Appendix \ref{app:algo} for discussion on boundary conditions). Indeed, we analyzed the same range of $\sigma$ and systems of the same size in the fully connected long-range model, using the traditional cluster algorithm of Luijten and B\"othe, and we verified that in the measure of $\eta$ comparable finite size effects are present. Angelini et al. \cite{angelini_relations_2014} indeed argue that near $\sigma = 1.75$ it is necessary to consider higher-order terms in the correction to scaling. In particular, they consider a scaling form of the type}
\begin{equation}
    \chi \sim L^{2-\eta}(a+bL^{-\delta}),
\label{eq:delta_correction}
\end{equation}
{where $a,b,\delta$ are parameters which in principle could depend on $\sigma$. They found that at $\sigma=1.75$ a value of $\delta \approx .42$ accounts for the correct scaling correction and the analytical predictions of Sak´s renormalization group are verified.  In Appendix \ref{app:peaks} we show indeed that a correction to the scaling with $\delta=.42$ is consistent with our simulation even for the other values of $\sigma$. In particular, Figure \ref{fig:eta_2d} shows that taking into account this correction to the scaling we obtain a very good agreement with Sak's predictions in the whole range of  $\sigma$.
Interestingly, in Appendix \ref{app:peaks} we also show that the ratio $a/b$ between the coefficients defined by Eq. \eqref{eq:delta_correction} strongly
depends on the exponent $\sigma$ and it displays a sharp minimum at $\sigma=1.75$. The presence of such a minimum is an evidence that finite size effects are particularly relevant at the transition between the SR and the LR regimes.} 
Our results are summarized in Tables \ref{tab:2d}, \ref{tab:2d_correction} and Figure \ref{fig:eta_2d}. More details about the numerical analysis  {and error estimate} can be found in the Appendix \ref{app:peaks}.

{We observe that the DLL, due to its locality, at criticality cannot be as efficient as the cluster algorithm in \cite{luijten_monte_1995} which involves non-local Montecarlo moves. Here, however, our goal was to show that the algorithm provides the expected equilibrium critical properties. In the next section, we use our local algorithm to study the dynamical critical properties that cannot be approached by non-local algorithms.}

\section{\label{sec:zeta} The dynamical critical exponent}

\subsection{Scaling analysis}

We now turn our attention to long range dynamical properties at criticality, a regime inaccessible to both cluster algorithms and kinetic Monte Carlo approaches. We concentrate on the dynamical critical exponent $z$, which governs the relaxation time. Indeed, near the critical point the autocorrelation time $\tau$, which is the Monte Carlo time needed to get two statistically independent samples, diverges as a power law $\tau \sim \xi^z$, where $\xi$ is the correlation length. When dealing with a critical finite size system, the only characteristic length scale is its linear size $L$, which implies
\begin{equation}
\tau \sim L^z.
\label{eq:tau_powerlaw}
\end{equation}
In this perspective we compute $\tau$ from the autocorrelation time of the absolute magnetization per spin.

We expect $z$ to show different values in the three regimes considered in equilibrium, as outlined in the Sak's scenario. In the MF region, the system behaves as a Curie-Weiss model. The underlying geometry disappears and the only remaining information is the total number of spins, i.e. the volume $V$. A simple argument shows that for the Curie-Weiss model, the relaxation time goes as $\tau \sim V^{1/2}$. Let us consider the Fokker-Planck equation for the magnetization:
\begin{eqnarray}
\frac{\partial P(m,\tau)}{\partial \tau} &=& \frac{\partial}{\partial m}\Big[ \,\Big(A(T-T_c)m + B m^3\Big)P(m,\tau)\Big] \,\nonumber \\
& & +\frac{\tilde{D}}{V}\frac{\partial^2}{\partial m^2}P(m,\tau),
\end{eqnarray}
where $A$ and $B$ are arbitrary constant and  the diffusion coefficient depends on the system size as $\frac{\tilde D}{V}$. Then, at criticality ($T=T_c$), the equilibrium distribution has the form $P^{(eq)}(m)=B e^{-\frac{C}{V}m^4}$, which implies $\langle m^2 \rangle \sim V^{-1/2}$. Since at criticality the magnetization remains small and the dynamics is expected to be purely diffusive, if we initialize the system at $m=0$ we get that $\langle m^2 \rangle \approx \frac{D}{V}t$. Thus we can conclude that the time needed to reach the equilibrium is $t\sim V^{1/2}$. Consequently, in the MF region we expect $\tau \sim L^{d/2}$.
At the crossing point $\sigma = d/2$, Renormalization Group calculations show that the model is Gaussian and the system is in a free super-diffusive regime in which the dynamical exponent $z$ is expected to be
\begin{equation}
    z = \frac{2d}{d_s},
\label{eq:z_diffusive}
\end{equation}
where $d_s$ is the spectral dimension, as defined in \cite{alexander_density_1982}.
In long range systems the spectral dimension is exactly $d_s^{(LR)} = 2d/\sigma$ \cite{burioni_geometrical_1997}, which means $z=\sigma$.
As soon as $\sigma>d/2$ we expect the free random walk behavior $z=\sigma$ to be perturbed by the presence of interactions, leading to a slightly larger value of $z$. This is consistent with the behavior of short range models, in which a small perturbation to the free diffusion is observed. In $d=1$ such a perturbed diffusive behavior should characterize the critical dynamics up to $\sigma=1$ where critical dynamics disappears. In $d=2,3$ we know that for the equilibrium critical exponents the short range behavior is recovered at $\sigma=2-\eta_{\rm SR}$. On the other hand, the non interacting dynamics of free random walks shows anomalous diffusion $z=\sigma$ up to $\sigma=2$ where normal diffusion ($z=2$) is recovered. In this perspective, for the dynamical exponent $z$ is not clear if the short range behavior is recovered at $\sigma=2-\eta_{\rm SR}$ or at $\sigma=2$. We show that our simulations seem to support the second hypothesis.

\subsection{The numerical measure of the autocorrelation time}

Let us call $m_i$ ($i=1\dots N$) the time series of the magnetization at a Monte Carlo step.
The autocorrelation time $\tau$ of $m_i$ is related to the error $\mathcal{E}$ on the mean $\overline{m}$ by the following relation \cite{sokal_monte_1997}:
\begin{equation}
\frac{\mathcal{E}^2}{\mathcal{E}_1^2} = \tau,
\label{eq:sigma_over_sigmazero}
\end{equation}
where $\mathcal{E}_1^2 = \frac{\sum_{i=1}^N (m_i - \overline{m})^2}{N(N-1)}$ is the estimate of the  error on $\overline{m}$ as if the system was uncorrelated.
The error $\mathcal{E}$ can be estimated by using the the Jackknife resampling method. We construct new samples aggregating bigger and bigger temporally consecutive blocks of magnetization. The aggregation is performed by taking the mean value of the block and then we calculate the error of the new sample, i.e.
\begin{equation}
\mathcal{E}(t_{BS})^2 = \frac{\sum_{i = 1}^{N/t_{BS}}(m^{BS}_i-\langle m \rangle)^2}{N/t_{BS}(N/t_{BS}-1)},
\label{eq:error_jacknife}
\end{equation}
where $t_{BS}$ indicates the block size and $m^{BS}_i$ is the mean value of the $i$-th block of aggregated magnetization. At large enough block sizes $t_{BS}$, $\mathcal{E}(t_{BS})^2$ turns out to be independent of $t_{BS}$ and the limit value of $\mathcal{E}(t_{BS})$ is the best estimator of the error on the mean of the time series.
At criticality in a finite system we expect the model to scale with the only intrinsic time in the dynamical evolution i.e. $L^z$. Since the block size of the Jackknife procedure is a time length, that we introduce to probe the system, we obtain:
\begin{equation}
\mathcal{E}_{N}(t_{BS},L) \equiv \frac{\mathcal{E}(t_{BS},L)}{\mathcal{E}(1,L)} = L^{z/2}\tilde{\mathcal{E}}_c(\sqrt{t_{BS}L^{-z}}),
\label{eq:rescaled_sigma}
\end{equation}
where $\tilde{\mathcal{E}}_c$ is a scaling function. 
Notice that for $t_{BS} \gg \tau$ we have $\mathcal{E}_{N}(t_{BS},L)= \tau^{1/2} \sim L^{z/2}$, so that  $\tilde{\mathcal{E}}_c(x)$ is constant in the limit of large $x$.
On the other hand, for $t_{BS} \ll \tau$ the 
variance of the resampled system in blocks $t_{BS}$ 
should be equal to the variance of the original time series. Therefore from Eq. (\ref{eq:error_jacknife},\ref{eq:rescaled_sigma}) we obtain $\tilde{\mathcal{E}}_c(x)\sim{x}$ for $x\sim 0$.
Eq. (\ref{eq:rescaled_sigma}) has been obtained exactly at $T=T_c$. Close to criticality we expect a general scaling relation $\mathcal{E}_{N}(t_{BS},t,L)= L^{z/2}\tilde{\mathcal{E}}(\sqrt{t_{BS}L^{-z}},
t L^{1/\nu})$,
where $t$ is the reduced temperature.
Since our simulations have been performed for each size $L$ at a temperature corresponding to the peak of the susceptibility $\chi$, from Eq. \eqref{eq:scaling chi}, $t L^{1/\nu}=x_M$ is constant in the different simulations. So we obtain $\mathcal{E}_{N}(t_{BS},L)= L^{z/2}\tilde{\mathcal{E}}(\sqrt{t_{BS}L^{-z}},
x_M)$ where the new scaling function $\tilde{\mathcal{E}}(\sqrt{t_{BS}L^{-z}},
x_M)=\tilde{\mathcal{E}}(\sqrt{t_{BS}L^{-z}})$ displays the same asymptotic features of the critical scaling function $\tilde{\mathcal{E}}_c(\sqrt{t_{BS}L^{-z}})=\tilde{\mathcal{E}}(\sqrt{t_{BS}L^{-z}},0)$.

The numerical data are analyzed using the exponent $z$ that gives the best collapse at different sizes $L$ for the rescaled functions  $L^{-z/2}\mathcal{E}_{N}$ as a function of $\sqrt{t_{BS}L^{-z}}$, see the Appendix \ref{app:zeta}  {for details and error estimate}.

\begin{figure}
\centering

\includegraphics[scale = 0.82]{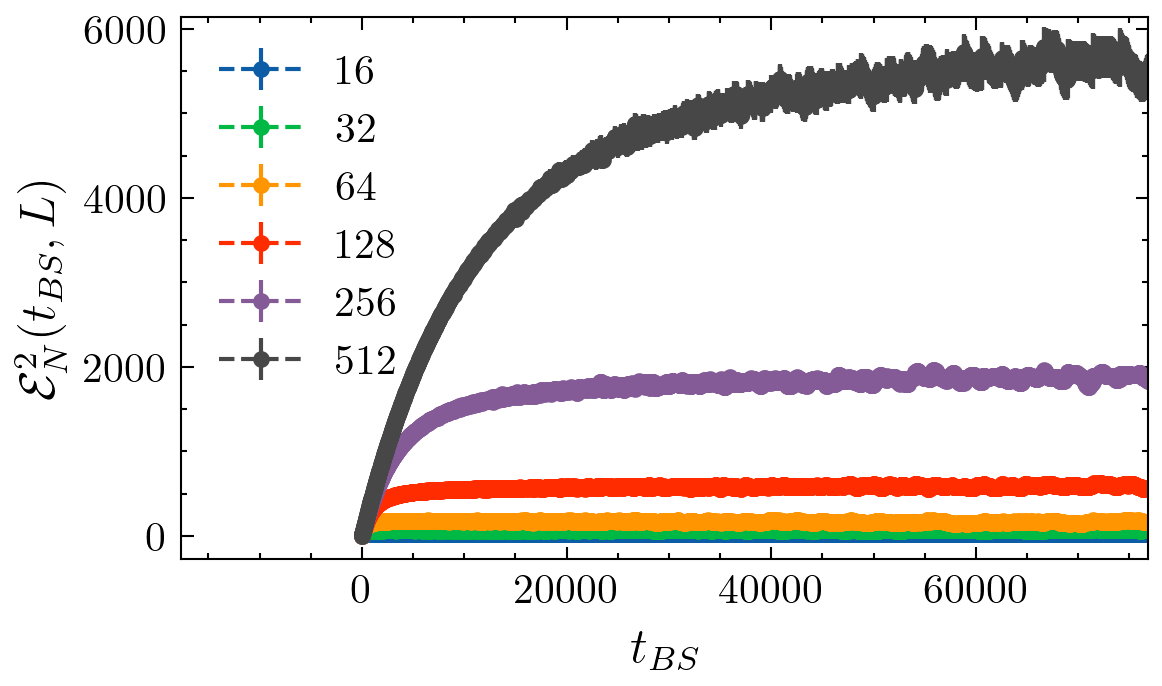}

\includegraphics[scale = 0.8]{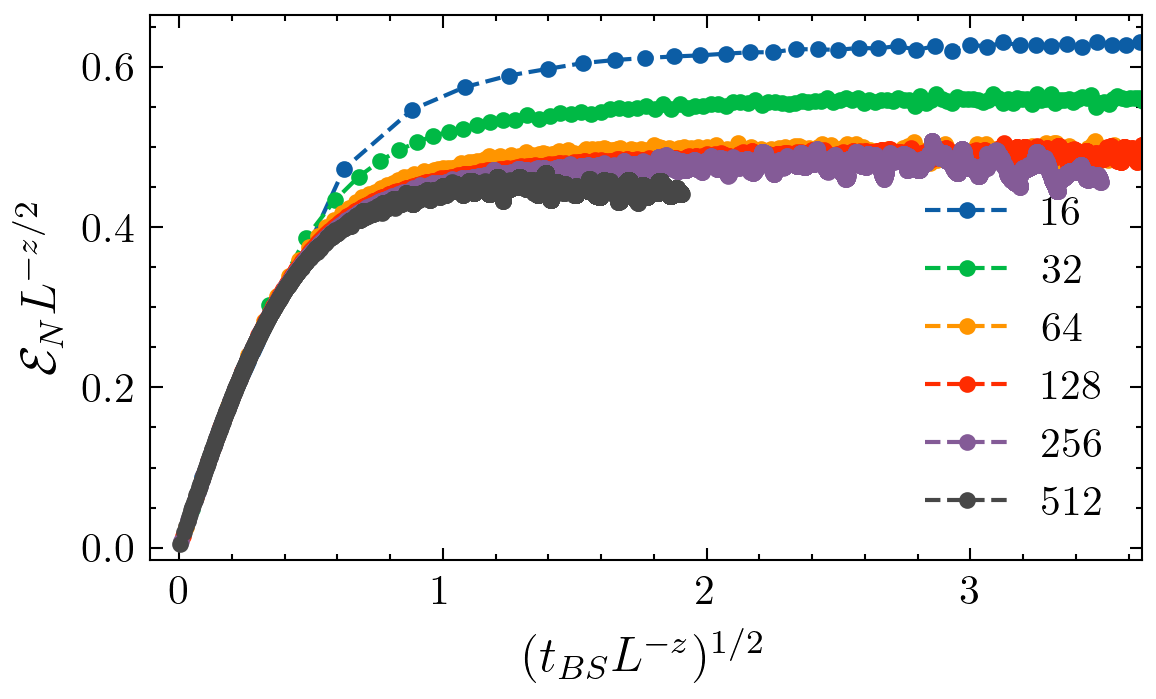}

\caption{Upper: $\mathcal{E}^2_N$ for $\sigma = 1.75$. The value of the autocorrelation time is the plateau. Lower: Rescaled $\mathcal{E}_N$  with $z = 1.75$, the collapse improves  with increasing size. The differences at small sizes are due to finite size effects.}
\label{fig:tau_J}
\end{figure}

In Figure \ref{fig:tau_J} we show an example with $\sigma = 1.75$, before and after rescaling. In upper panel, we clearly see the effect of critical slowing down, by looking to the fast increasing plateau with the size. The rescaling is performed with the best value of $z$, showing a collapse which improves with increasing system size.

{
The method we introduced turns out to be quite efficient. In fact, for systems with strong finite size effects, a good estimate of $z$ requires including in the scaling analysis even sizes where the simulation times are not much larger than the decorrelation time. This can be observed in figure \ref{fig:tau_J} for $L=512$, where a plateau is not yet reached at large block sizes $t_{BS}$ and the decorrelation time cannot be directly calculated by Eq. \eqref{eq:sigma_over_sigmazero} by measuring a stable asymptotic value. Nevertheless, with our method the system at $L=512$ can be included in finite-dimensional rescaling, providing an important contribution to the estimate of $z$. An alternative way to measure the $z$ exponents using even shorter simulations is through 
a critical quench \cite{Hasenbusch}, in which equilibration of the system is not even required. This method, however, implies knowledge of the static equilibrium critical exponents.}

\subsection{Results in $1$ and $2$ dimension}

\begin{figure}
    \centering
    \includegraphics[scale = 0.8]{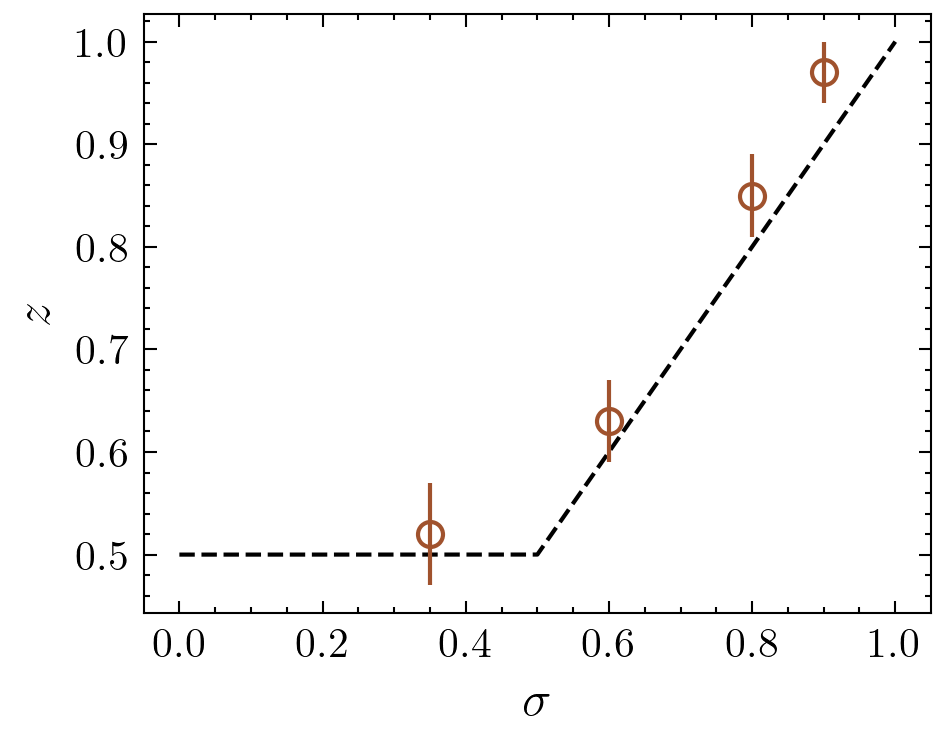}
    \caption{$z$ The dynamical critical exponent as obtained from DLL (see Appendix \ref{app:zeta}) for the $1$-dimensional case. The black dashed line represents the theoretical prediction at leading order, as discussed in the text. Numerical data are illustrated in Table (\ref{tab:zeta_1d}) in Appendix \ref{app:tables}}
    \label{fig:zeta_1d}
\end{figure}

\begin{figure}
    \centering
    \includegraphics[scale = 0.8]{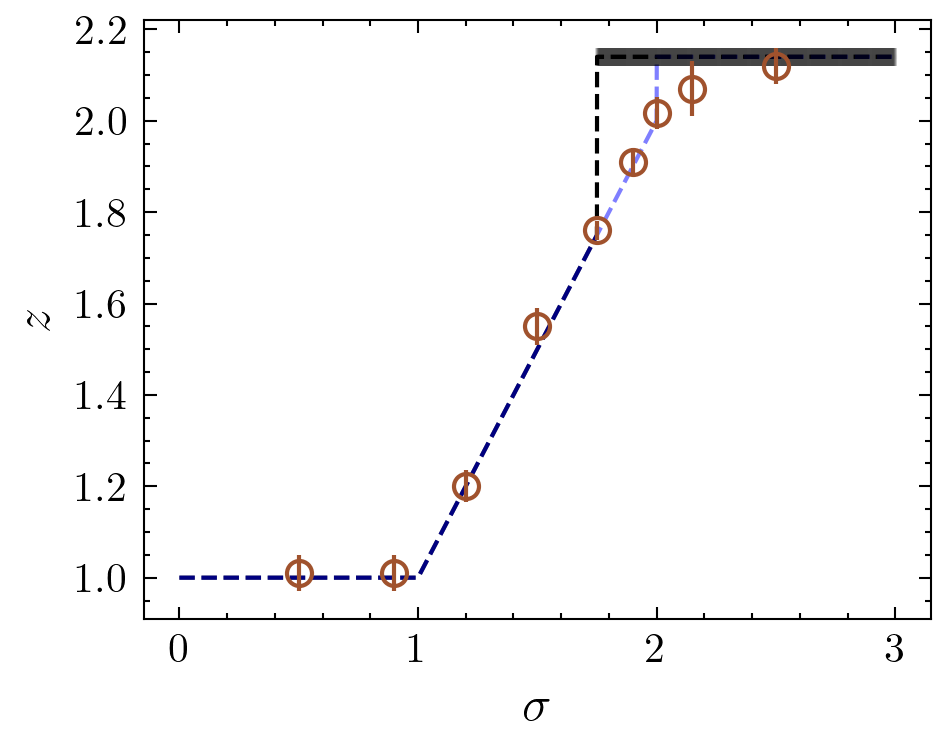}
    \caption{The dynamical critical exponent as obtained from DLL (see Appendix \ref{app:zeta}), for the $2$-dimensional case. Again, the black dashed line represents the leading order, following the Sak's static prediction. The lighter dashed line represents the leading order in the case where the SR behaviour is recovered at $\sigma = 2$. The dark horizontal line at higher values of $\sigma$ is the value of $z$ with its error as found in \cite{adzhemyan_dynamic_2022}, $z = 2.14(2)$. Numerical data are illustrated in Table (\ref{tab:zeta_2d}) in Appendix \ref{app:tables}.}
    \label{fig:zeta_2d}
\end{figure}

We collect the results for both $1$ and $2$-dimensional case in Tables \ref{tab:zeta_1d} and \ref{tab:zeta_2d}, and in Figures \ref{fig:zeta_1d} and \ref{fig:zeta_2d}. We recall that, in the $1$-dimensional case, there is no phase transition in the SR region. In the $2$-dimensional case for this region we compare our results with $z = 2.14 \pm 0.02$, as taken from \cite{adzhemyan_dynamic_2022}.

The values obtained for the $1$-dimensional case are consistent with our analysis. For the more interesting $2$-dimensional case, in Figure \ref{fig:zeta_2d} the darker line represents the short range behavior according the Sak's hypothesis on the onset of LR and SR regimes. If $\sigma \in (1.75, 2)$ corresponds to SR region, we should observe a rapid deviation from the free super-diffusive behavior $z=\sigma$ so that the short range behavior should be recovered at $\sigma = 1.75$.  Instead we found that it seems to be recovered smoothly at $\sigma=2$. In this perspective, in order to obtain a clear picture on the transition in the dynamical exponent from the LR to the SR regime, more extensive simulations are required to clarify the picture in the region $\sigma \in (1.75, 2)$, together with some analytic Renormalization Group argument which is beyond the scope of our work.


\section{\label{sec:conclusion}Conclusion and perspectives}

In this work we have presented an analysis of the equilibrium critical exponents and of the dynamical exponent $z$ at criticality for $1$ and $2$-dimensional Ising models with long range interactions. The numerical values of the exponents have been obtained using a new dynamical algorithm, the \textit{Dynamical Levy Lattice}, designed for the study of models with long range interactions. The algorithm corresponds, in practice, to a Monte Carlo analysis on a diluted temporal graph.
The dynamics has relatively small computational cost, avoiding the typical scalability problems of long range interactions, preserves locality, and can be adapted to a wide range of situations beyond Ising interactions. However, since detailed balance is not satisfied and the stationary distribution is not known, the equivalence between our model and the long range Ising equilibrium distribution has been checked. Our results are in line with the present literature.  
Exploiting the local nature of our algorithm, we then obtain a first measure of the dynamical critical exponent $z$, which shows a peculiar behaviour near the LR-SR crossing point. Interestingly, the SR regime seems to be reached at larger values of $\sigma$ than in the equilibrium case.  Extensive simulations and new analytical arguments are needed to clarify this scenario. 
The effect of a different choice of $q$ on the dynamics is still to be carefully tested. Our choice $q = 3$ is driven by the results of the fully connected model, in which analytical calculations can be performed exactly.
\cite{jedrzejewski_oscillating_2015,chmiel_q-neighbor_2018} Our choice gives a quite effective dynamics, however more efficient choices could be possible. 
The locality of the algorithm makes it widely applicable to study out-of-equilibrium properties, for example to investigate aging in critical quenches or the effect of long range interaction in coarsening phenomena \cite{christiansen_aging_2020}, in which topology is expected to have non trivial effects such as pinning \cite{Corberi1,burioni_phase_ordering_2007,burioni_topological_2013,corberi_phase_ordering_2015}.  
Finally, the numerical approach can be easily extended to different models. In particular, the XY model in $d=2$ is expected to show non trivial features in the presence long range interactions, due to the three different phases, i.e. paramagnetic, ferromagnetic and Kosterlitz Thouless \cite{giachetti_berezinskii-kosterlitz-thouless_2021}.

\begin{acknowledgments}
We warmly thank Maria Chiara Angelini and Federico Corberi for very useful discussions and suggestions. We thank Matteo Ghizzi for providing part of the numerical data for the $1$-dimensional equilibrium case. 
This research benefits from the HPC (High Performance Computing) facility of the University of Parma.
\end{acknowledgments}

\appendix

\section{\label{app:algo}Algorithm implementation}
In the algorithm of the DLL, we start by choosing uniformly random a spin $\sigma_i$. Next we choose a list $S_q$ of $q$ spins, which will be neighbours of $i$ during this temporal step. Such $q$ spins $\sigma_j$ are drawn from the distribution $P(r_{ij}) \propto r_{ij}^{-(d+\sigma)}$, with $r_{ij}$ being the distance between nodes $i$ and $j$ and $\sigma$ the decay parameter. A precise definition of $P(r_{ij})$ requires the introduction of boundary conditions, that will be discussed later. Chosen $i$ and its $q$ neighbors, we apply the standard dynamics, for example Metropolis or Glauber, using as flipping energy $\Delta E = 2\cdot J\cdot \sigma_i\cdot \sum_{j\in S_q} \sigma_{j}$. Applying $N$ times this procedure defines a Monte Carlo step. The procedure is summarized in Algorithm \ref{alg:cap}.
\begin{algorithm}[H]
\caption{DLL (single Monte Carlo step)}\label{alg:cap}
\begin{algorithmic}[1]
\Require $q \in \mathbb{N}$ and $\sigma \in \mathbb{R}^+$
\State Choose a random spin $\sigma_i$
\State Draw a node  $j\not =i$ from the probability distribution $P(r_{ij}) \sim r_{i,j}^{-(d+\sigma)}$ where $r_{ij}$ is the distance between nodes $i$ and $j$. Repeat the extraction $q$ times so that you get $q$ random nodes, (allow for the extraction of the same node more than one time in the list $S_q$).  
\State Calculate the interaction energy $\Delta E = 2\cdot J\cdot \sigma_{i}\cdot \sum_{j\in S_q} \sigma_{j}$
\State Flip $\sigma_i$ following a dynamics, Metropolis or Glauber, using $\Delta E$ as flipping energy
\State Repeat steps 1-5 $O(N)$ times
\end{algorithmic}
\end{algorithm}
In the fully connected standard q-Ising model i.e. $P(r_{ij})=constant$, analytical calculations show that for $q<2$ the system does not present a phase transition at finite temperature i.e. $T_c=0$; with Metropolis dynamics, for $q=3$ there is a continuous phase transition of the mean field universality class while for $q>2$ a first order transition is observed \cite{jedrzejewski_oscillating_2015,chmiel_q-neighbor_2018}. For Glauber dynamics instead, a continuous transition is always observed for $q>2$. For this reason we choose the Glauber dynamics and set $q=3$ so that we expect a second order phase transition, as we indeed observe in our simulations. In particular, we verify that both lowering and increasing temperature the system never presents hysteresis. 
A small value of $q$ allows for an efficient implementation of the algorithm while large values are typically more demanding since the algorithm requires the extraction of $q$ random numbers in a microscopic step. In general the study of the dependence on $q$ of the dynamical evolution is an interesting open issue.
Let us briefly discuss the choice of the boundary conditions. As it is well known for the fully connected long range model the results at small $\sigma$ are strongly affected by finite size effects and by the choice of boundary conditions. The most natural choice is to  extract as a distance $r_{ij}$ a $d$-dimensional integer vector, from the probability distribution $p(\vec{r}_{ij})\sim |\vec{r}_{ij}|^{-(d+\sigma)}$ with $\sqrt{d}/2<| \vec {r}_{ij}|<L/2$ and then impose periodic boundary conditions. In \cite{angelini_relations_2014} it is shown that a more efficient choice of the boundary condition is to use copies (images) of the original configuration; in our DLL this means to drawn the integer vector $\vec{r}_{ij}$ from the probability distribution $p(\vec{r}_{ij}) \sim |\vec{r}_{ij}|^{-(d+\sigma)}$ with $\sqrt{d}/2<| \vec {r}_{ij}|$ (no upper limit to the distribution). The site $j$ is then obtained with periodic boundary condition where now $\vec{r}_{ij}$ can wind around the torus an arbitrary number of times. We verify indeed that also for our model this choice is more efficient in simulations. A further choice of the boundary conditions turns out to be even slightly better. In particular, we draw again the vector $\vec{r}_{ij}$ from the integer probability distribution $p(\vec{r}_{ij}) \sim |\vec{r}_{ij}|^{-(d+\sigma)}$ with $\sqrt{d}/2<| \vec {r}_{ij}|$ (no upper limit); then we fix $\sigma_j=0$ if $| \vec {r}_{ij}|>L/2$; while for $| \vec {r}|<L/2$ the node $j$ is obtained imposing periodic boundary conditions; in this way when faraway coordinates are considered we take into account that interaction occurs with nodes of zero magnetization on average, while spin-spin correlations are relevant only at distances smaller than $L$. In $d=2$ finite size effects are large and we adopt this last prescription for the boundary conditions, which better reproduces Sak predictions on the exponent $\eta$.  On the other hand, in $d=1$ the results seem independent of the choice since we deal with very large $L$.

\section{\label{app:peaks}Simulation details - the susceptibility}

\begin{figure}
    \includegraphics[scale = 0.99]{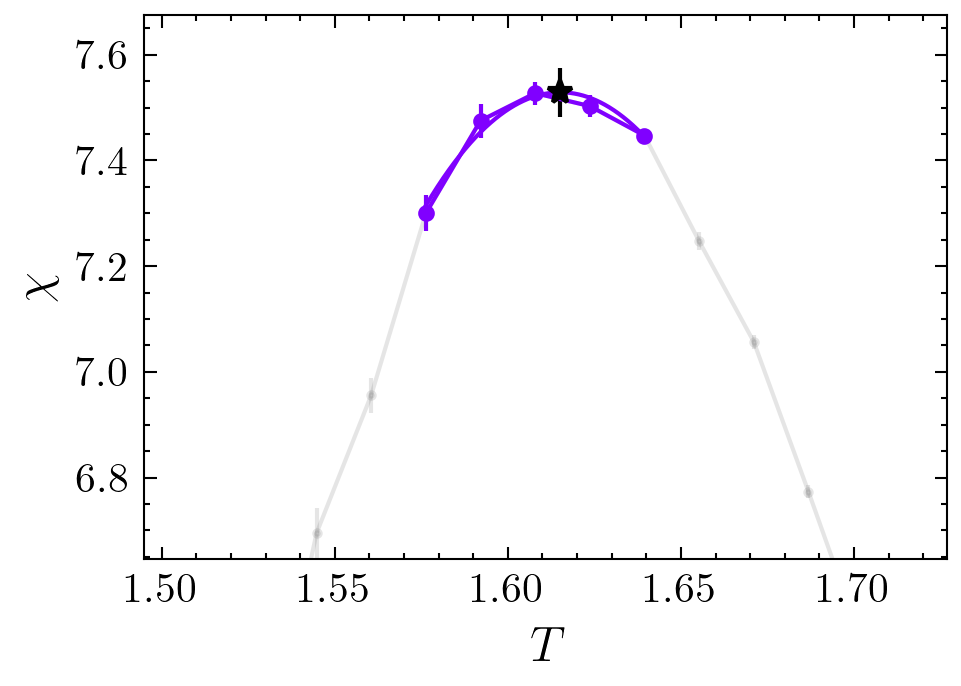}
    \caption{Parable extrapolation of the maximum of $\chi$ for $d=2$, $\sigma = 1.2$ and $L=16$, which result is drawn in the same color of the points. The star is the extrapolated maximum of the curve, with its respective error bar.}
    \label{fig:parable}
\end{figure}

During the Monte Carlo simulation we save the value of the magnetization $m = L^{-d}\sum_{i=1}^N \sigma_i$, for a total of $O(10^6)$ realization for each temperature. Then we compute the susceptibility as 
\begin{equation}
    \chi = k_bTL^{-d}(\langle m^2\rangle- \langle |m|\rangle^2
    \label{eq:chi}
\end{equation}
where the constant $k_b$ is set as equal to $1$. The error for each $\chi$ is calculated taking the standard deviation of different Markov chain realizations. {These errors are propagated in every successive fit we make, always checking that the reduced $\chi^2$ is compatible with the unity, for each one of those.} 
To find the maximum value of $\chi$, we concentrate our simulations near the peak and then we extract the maximum with a quadratic fit{, propagating the errors as mentioned before}. A sketch of this procedure can be seen at Figure \ref{fig:parable}. 
Therefore, we perform a linear fit with the logarithm of the maximum,
in particular from equation (\ref{eq:scaling chi}) we get:
\begin{equation}
    \label{eq:chi fit}
    \log{(\chi_{max}(L))} = y\cdot\log{L}+C,
\end{equation}
with $\chi_{max}$ the maximum of $\chi$ and C a constant. 

\begin{figure}
    \centering
    \includegraphics[scale = 0.9]{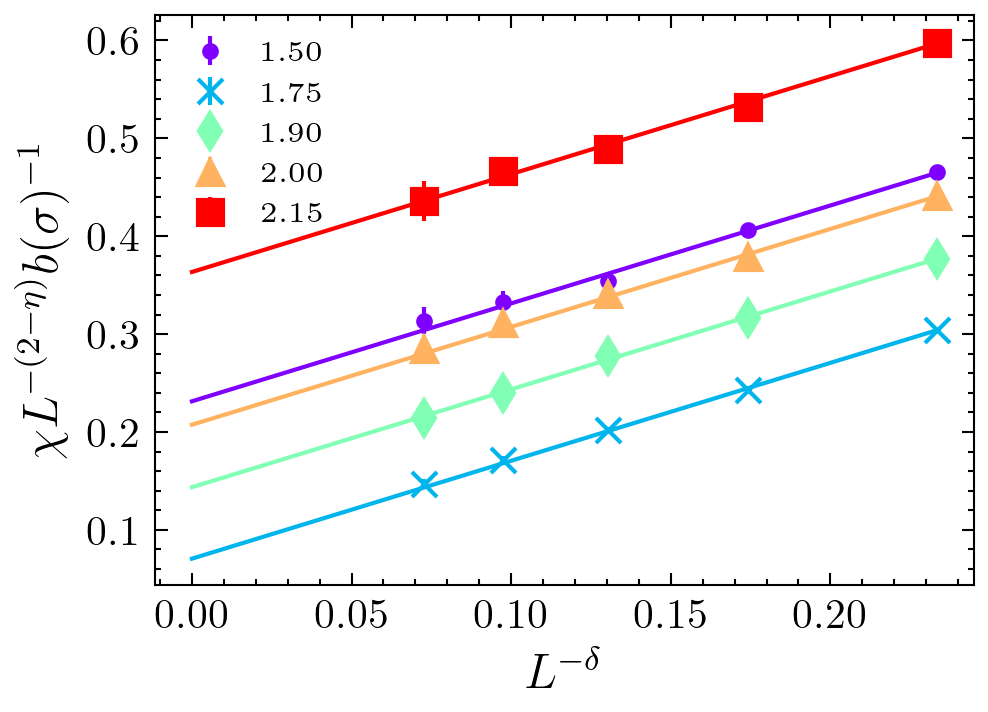}
    \caption{Linear dependence of $\chi L^{-(2-\eta)}$ for $\sigma$ values near the SR-LR crossing point.Straight line are the result of the linear fit respect to $L^{-\delta}$}. 
    \label{fig:linear_correction}
\end{figure}

\begin{figure}
    \centering
    \includegraphics[scale = 0.9]{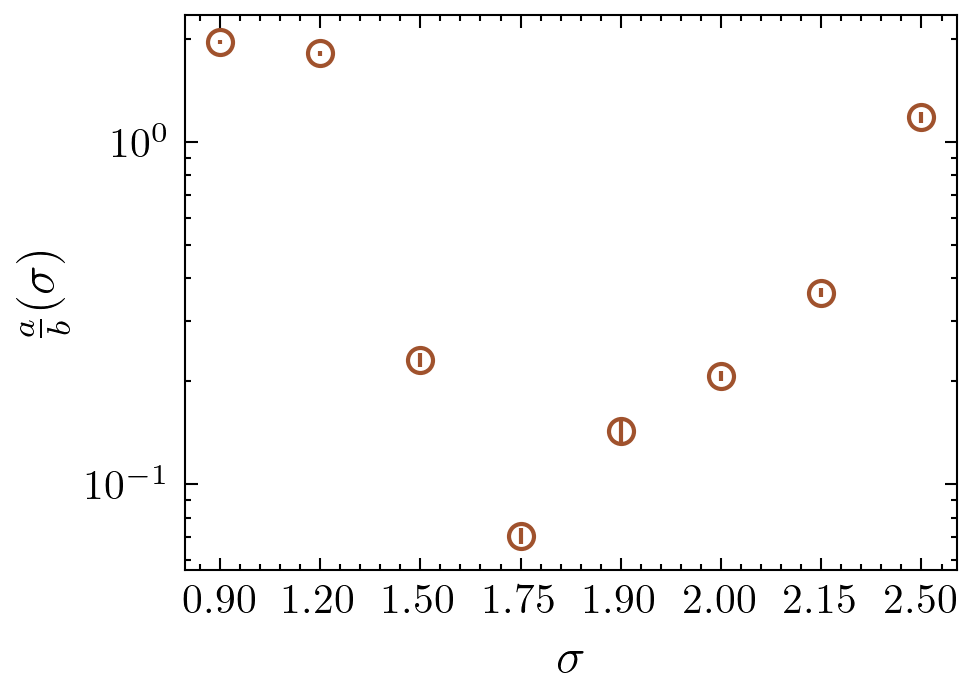}
    \caption{$\frac{a}{b}$ ratio for different values of $\sigma$. The logarithmic scale on the $y$ axis, shows how fast this ratio grows as we move away from the SR-LR crossing point.}
    \label{fig:a_over_b}
\end{figure}

{
Typically Eq. \eqref{eq:chi fit} is affected by strong finite size effects and the linear behavior in log-log plot is not observed. Thus our strategy is to measure the variation of the slope $y$ 
as a function of the size $L$ as illustrated in Figure \ref{fig:fss_sketch}. For such size dependent slopes $y(L)$ we extrapolate the value at $L=\infty$ by means of a linear fit against $L^{-1}$, i.e. we assume the intercept of this fit as an estimate of the exponent. Again, at this step we propagate the errors coming from the quadratic fits.
The result of this simple extrapolation procedure is shown in the circle of Figure \ref{fig:eta_2d} evidencing that still important finite size effects are present in the estimate. Indeed, in \cite{angelini_relations_2014} it has been shown that the correction to the scaling at $\sigma=1.75$ vanishes more slowly  than linearly according to Eq. \eqref{eq:delta_correction}. In this perspective we extrapolate the value of $y$ by fitting linearly the size dependent estimates $y(L)$ as a function of $(1/L)^{\delta}$ and we obtain the data shown with blue stars in Figure \ref{fig:eta_2d} which confirm Sak's prediction on the exponent $\eta$. In this second approach errors are larger (see Figure \ref{fig:eta_2d}) since it assumes larger scaling corrections.
\begin{figure}
    \centering
    \includegraphics[scale = 0.8]{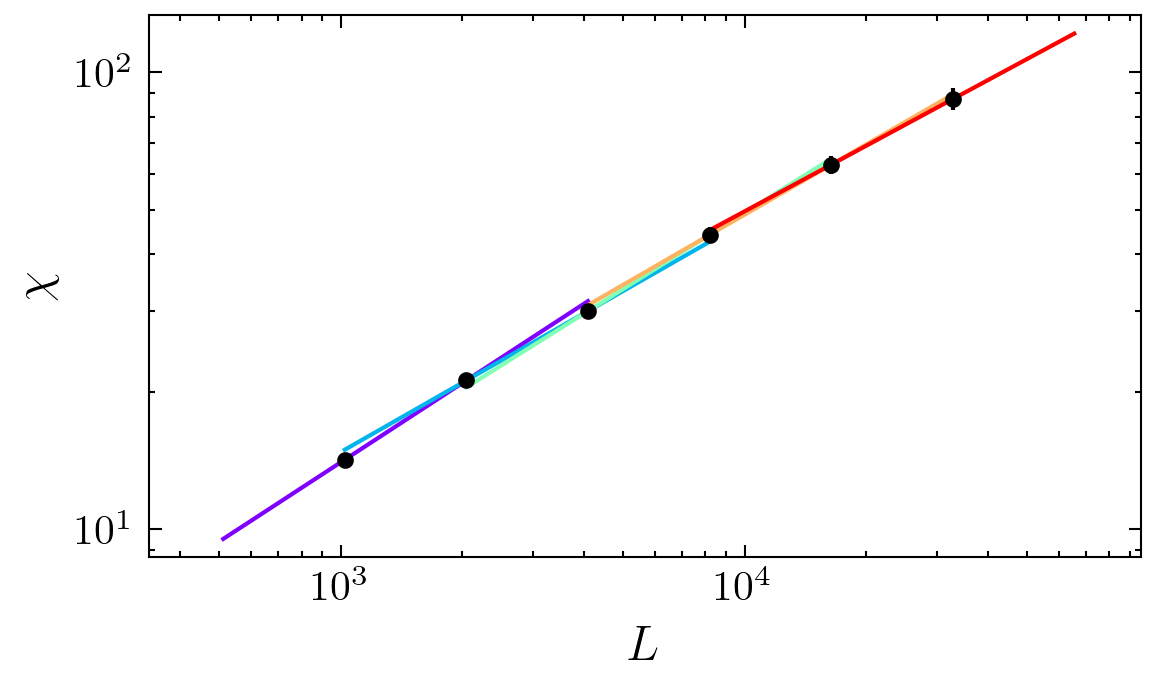}
    \caption{Finite size extrapolation example for $d = 1$ and $\sigma = 0.35$. Each line is the result of a linear fit of two consecutive sizes and it is clear how the slope changes with increasing sizes.}
    \label{fig:fss_sketch}
\end{figure}

Figure \ref{fig:linear_correction} shows that corrections to scaling according to Eq. \eqref{eq:delta_correction} are indeed consistent with our numerical simulations. In particular, 
the quantity $\chi L^{-(2-\eta)}$ displays a linear dependence respect to $L^{-\delta}$ with $\delta=0.42$. Then we study the ratio $\frac{a}{b}$ where $a$ and $b$ are the coefficient in Eq. \eqref{eq:delta_correction} obtained from the linear fit in Figure \ref{fig:linear_correction}. In Figure  \ref{fig:a_over_b} we find that $\frac{a}{b}$ has a sharp minimum for $\sigma = 1.75$ and it grows when moving away from the SR-LR crossing point. This confirms that the corrections to scaling are bigger in this transition regime. 
}

\section{\label{app:zeta}Simulation details - $z$ extrapolation}
In this section we briefly explain the method we used to calculate the $z$ dynamical critical exponent, since there is no standard way to measure it. We define a function which depends on $z$, whose minimum indicates the best estimate of the exponent. The function is defined as follows. First, we take a value of $z$ and we rescale $\mathcal{E}_N$, as mentioned previously in the text, i.e we look at the curves $L^{-z/2}\mathcal{E}_N(\tilde{t},L)$ as function of $\tilde{t} \equiv (t_{BS}L^{-z})^{1/2}$, at different sizes $L$. Then, for each value of $\tilde{t}$, we compute the difference between consecutive curves in size, defining the function:
\begin{equation}
\Delta\mathcal{E}_N(\tilde{t},L) \equiv \mathcal{E}_N(\tilde{t},L)L^{-z/2} - \mathcal{E}_N(\tilde{t},L/2)(L/2)^{-z/2}
\end{equation}
To take into account finite size effects, we extrapolate the thermodynamic limit of $\Delta\mathcal{E}_N(\tilde{t},L)$, by assuming a linear dependence of $L^{-1}$:
\begin{equation}
    \Delta\mathcal{E}_N(\tilde{t},L) = AL^{-1} + \Delta\mathcal{E}_N(\tilde{t},\infty).
\end{equation}

{In this case, a next to leading correction to the scaling e.g. as $L^{-\delta}$ with $\delta\not=1$, seems to have a small effect on the final value of $z$, at least for the size and errors we are considering in our simulation.} 
Finally, we take the square sum of $\Delta\mathcal{E}_N(\tilde{t},\infty)$ for all the values of $\tilde{t}$, to penalize the discrepancies from $0$. Indeed, the best value of $z$ is the one for which, in the thermodynamic limit, the curves at different size collapse. Thus, for each value of $z$  we have defined the function:
\begin{equation}
    \Delta\mathcal{E}_N(z) \equiv \sum_{\{\tilde{t}\}} \Delta\mathcal{E}_N^2(\tilde{t},\infty),
\end{equation}
where the sum is extended to all value of $\tilde{t}$. Our best estimate of $z$ is the one for which this function has a minimum. For the error, we look at the values of $z_e$ for which $\Delta\mathcal{E}_N(z_e) = 5\times \Delta\mathcal{E}_N(z_{min}) $.

\section{\label{app:tables}Numerical data}

\begin{table}[!h]
\begin{ruledtabular}
\begin{tabular}{ccc}

$\sigma$  & Experiment & $\Delta y$           \\ 
\colrule
$0.35$ & $0.507 \pm  0.015$ & $0.007 \pm 0.015$  \\ 
$0.6$ & $0.612  \pm  0.016$ & $0.012 \pm 0.016$  \\ 
$0.8$ & $0.761 \pm  0.032$ & $0.039 \pm 0.032$  \\ 
$0.9$   & $0.861 \pm  0.041$  & $0.039 \pm 0.041$  \\ 
\end{tabular}
\caption{$y$ exponent extrapolation for $1$-dimensional spin chain. The last column indicates the discrepancy with respect to the expected theoretical exponent.}
\label{tab:1d}
\end{ruledtabular}
\end{table}

\begin{table}[!h]
\begin{ruledtabular}
\begin{tabular}{ccc}

$\sigma$ & Experiment  & $\Delta y$   \\ 
\colrule
$0.5$ & $0.95 \pm 0.03 $ & $0.05 \pm 0.03$\\ 
 $0.9$ & $1.019 \pm 0.025 $ & $0.019 \pm 0.025$\\ 
 $1.2$ & $1.168 \pm 0.033 $ & $0.032 \pm 0.033$\\ 
 $1.5$ & $1.424 \pm 0.061 $ & $0.076 \pm 0.061$\\ 
 $1.75$ & $1.608 \pm 0.032 $ & $0.142 \pm 0.032$\\ 
 $1.9$ & $1.679 \pm 0.046 $ & $0.071 \pm 0.046$\\ 
 $2.0$ & $1.685 \pm 0.048 $ & $0.065 \pm 0.048$\\ 
 $2.15$ & $1.705 \pm 0.038 $ & $0.045 \pm 0.038$\\
 $2.5$ & $1.754 \pm 0.026 $ & $0.004 \pm 0.026$\\ 

\end{tabular}
\caption{$y$ exponent extrapolation for $2$-dimensional lattice. The last column indicates the discrepancy with respect to the expected theoretical exponent.}
\label{tab:2d}
\end{ruledtabular}
\end{table}

\begin{table}[!h]
\begin{ruledtabular}
\begin{tabular}{ccc}

$\sigma$ & Experiment  & $\Delta y$   \\ 
\colrule
 $0.5$ & $1.022 \pm 0.059 $ & $0.022 \pm 0.059$\\ 
 $0.9$ & $0.995 \pm 0.051 $ & $0.005 \pm 0.051$\\ 
 $1.2$ & $1.169 \pm 0.064 $ & $0.031 \pm 0.064$\\ 
 $1.5$ & $1.516 \pm 0.062 $ & $0.016 \pm 0.062$\\ 
 $1.75$ & $1.766 \pm 0.068 $ & $0.016 \pm 0.068$\\ 
 $1.9$ & $1.766 \pm 0.069 $ & $0.016 \pm 0.069$\\ 
 $2.0$ & $1.761 \pm 0.07 $ & $0.011 \pm 0.07$\\ 
 $2.15$ & $1.793 \pm 0.076 $ & $0.043 \pm 0.076$\\ 
 $2.5$ & $1.789 \pm 0.075 $ & $0.039 \pm 0.075$\\ 
\end{tabular}
\caption{$y$ exponent extrapolation for $2$-dimensional lattice when using second order correction $\delta = 0.42$, as discussed in the text.}
\label{tab:2d_correction}
\end{ruledtabular}
\end{table}

\begin{table}[!h]
\begin{ruledtabular}
    
\begin{tabular}{cc}

$\sigma$ & $z$    \\ 
\colrule
 $0.35$ & $0.52 \pm 0.05 $ \\ 
  $0.6$ & $0.63 \pm 0.04 $ \\ 
$0.8$ & $0.85 \pm 0.04 $ \\ 
 $0.9$ & $0.97 \pm 0.03 $ \\

\end{tabular}
\caption{$z$ exponent extrapolation for the $1$-dimensional case.}
\label{tab:zeta_1d}
\end{ruledtabular}
\end{table}

\begin{table}[!h]
\begin{ruledtabular}
    
\begin{tabular}{cc}

$\sigma$ & $z$    \\ 
\colrule
 $0.5$ & $1.01 \pm 0.04 $ \\ 
 $0.9$ & $1.01 \pm 0.04 $ \\ 
 $1.2$ & $1.2 \pm 0.035 $ \\ 
 $1.5$ & $1.55 \pm 0.04 $ \\ 
 $1.75$ & $1.76 \pm 0.02 $ \\ 
 $1.9$ & $1.91 \pm 0.03 $ \\ 
 $2.0$ & $2.016 \pm 0.035 $ \\ 
 $2.15$ & $2.07 \pm 0.06 $ \\ 
 $2.5$ & $2.12 \pm 0.04 $ \\

\end{tabular}
\caption{$z$ exponent extrapolation for the $2$-dimensional case.}
\label{tab:zeta_2d}
\end{ruledtabular}
\end{table}

\bibliography{DLL}

\end{document}